 \documentstyle[preprint,aps]{revtex}
\begin{document}
\draft
\preprint{Submitted to Computers in Physics}
\title{Four-tap shift-register-sequence random-number generators}
\author{Robert M.  Ziff}
\address{
Department of Chemical Engineering,
University of Michigan,
Ann Arbor, MI 48109-2136
}
\date{\today}
\maketitle
\begin{abstract} 
It is shown how  
correlations in the generalized feedback shift-register (GFSR)
random-number generator are greatly diminished when the 
number of  feedback taps is increased from two to four 
(or more) and the tap offsets are lengthened.
Simple formulas for 
producing maximal-cycle four-tap rules from available 
primitive trinomials are given, and explicit
three- and four-point correlations are found for some
of those rules.
A number of generators are also tested using a simple
but sensitive random-walk simulation that relates to a problem in 
percolation theory.
While virtually all two-tap generators fail this test,
four-tap generators
with offset greater than about 500 pass it, have passed tests carried
out by others,
and appear to be good multi-purpose high-quality random-number generators.
\end{abstract}
\pacs{}

\narrowtext

\section{Introduction}

The generalized feedback shift-register (GFSR) random-number
generator R$(a,b,c,\ldots)$ produces pseudo-random numbers by
the linear recursion \cite{taus65,gol67,lewis73}
\begin{equation}
x_n = x_{n-a}\ \oplus \ x_{n-b}\ \oplus \ x_{n-c}  \oplus \ \ldots 
\label{eq:recur}
\end{equation}
where $\oplus$ is the exclusive-or operation (addition 
modulo 2) and $a, b, c, \ldots$ are the feedback taps.
Here the $x_n$ are either single bits or multi-bit words,
in which case the $\oplus$ operation is carried 
out bit-wise.
This recursion was first studied extensively by 
Golomb \cite{gol67} in the context
of computer science, where it has many other applications,
including cryptology and error-correcting codes. 
Its use as a random-number generator was introduced to the
computational physics community by Kirkpatrick and Stoll 
\cite{kirk81}, who suggested the two-tap rule R$(103, 250)$, 
and became fairly popular due to its simplicity and
generally accepted quality.

However, it is now widely known that such generators,
in particular with the two-tap rules such as
R$(103,250)$, have serious deficiencies.
Many years ago, Compagner and Hoogland \cite{comp87} reported 
irregularities in an Ising model simulation using R$(15, 127)$. 
The present author found problems using  R$(103, 250)$ in a hull-walk
simulation \cite{ziff86}, and switched to an empirical
combination generator\cite{ziff87}.
Marsaglia \cite{mars85} observed very poor behavior with 
 R(24,55) and smaller generators, and advised 
against using generators of this type altogether.

More recently, Ferrenberg et al.~\cite{ferr92} 
found that R$(103,250)$ leads to results
being more than 100 standard deviations from the (known) true
values, in simulations of the Ising 
model with the Wolff cluster-flipping Monte-Carlo algorithm. 
Coddington \cite{codd} confirmed this observation with an extensive study
involving a large number of various random-number generators.
Grassberger found striking errors in an 
efficient depth-first self-avoiding random-walk algorithm
when R$(103,250)$ was used \cite{grass94}. 
Vattulainen et al.~\cite{vattu94} devised a number of simple
tests that clearly show the effective
correlations and deficiencies in two-tap GFSR generators.
And very recently, Shchur et al.
\cite{shchur} simplified the one-dimensional Wolff algorithm to a
repeating one-dimensional
random walk test, which they showed fails dramatically when
R$(103,250)$ is used.

The basic problem of two-tap generators R$(a,b)$ 
 is that they have a built-in
three-point correlation between $x_n$, $x_{n-a}$, and $x_{n-b}$,
simply given by the generator itself,
such that if any two of
the $x_n$ are known, the third follows directly from the
recursion $x_n = x_{n-a}\ \oplus \ x_{n-b}$. 
 While
these correlations are spread over the size
$p =$ max$(a,b,c,\ldots)$ of the generator itself,
they can evidently still lead to significant errors.
These three-point correlations have been recently brought out clearly
in a simulation by Schmid and Wilding \cite{schmid}

Other problems with this generator are also known.
Compagner and Hoogland \cite{comp87}  have shown how a pattern
of all 1's in the initialization string leads to complex
(and beautiful) pattern of subsequent bits that persists for a
surprisingly long time.  Shchur et al.\ \cite{shchur}
showed that, if an event occurs
with a probability close to one (such as 31/32), it is not too unlikely
for say 249 successive true outcomes to occur, which then
leads to a very
serious error at the 250-th step, when the R$(103,250)$ generator is used.

For reasons like these, many people have, over the years, 
advocated using larger tap offset values $a, b, \ldots$ and
increasing the number of those taps from two to four or more (they
are always even in number for maximal-cycle generators).
Compagner and co-workers have considered generators with offsets as
large as 132049 \cite{her92}, and have proposed combining
two generators to effectively make
multi-tap rules, which possess good behavior
\cite{comp91,comp91a}.  The advantages of using larger offsets are
well documented; for example, Ferrenberg et al.\ found the 
generator R$(216,1279)$ to be nearly acceptable for their problem,
and Coddington showed that 
R$(1393,4423)$ reduces the error below the measurable limit
for the simulation cutoff that he used.  Similar trends were seen by
by Compagner and Hoogland \cite{comp87} and
Vattulainen et al.\ {\cite{vattu94}}.

However, the use of more 
than two taps has not been common in practice. One 
reason is undoubtedly that tables of primitive polynomials on GF(2)
(the Galois field on binary numbers)
 of order higher than three, which are needed to construct 
maximal-cycle rules, have been limited (although some have appeared 
more recently \cite{kurita,zivkovic}), and their direct 
determination is a non-trivial exercise in number theory.
Golomb has given a prescription for making new 
generators from existing ones based upon sequence decimation 
\cite{gol69}, which can be used to construct multi-tap rules.
In the present paper, I 
simplify this procedure by giving explicit formulas
for 3-, 5-, and 7-decimation
of two-tap rules, in which cases
four-tap rules always result. 
These four-tap rules generate, in single
calls, the same sequences that come from $D$-decimation of
the the two-tap generators they derive from.

It turns out that this decimation procedure has been frequently 
employed in a literal sense: simply by using every $D$-th call
of a given generator. For example, Ferrenberg
et al.~considered using every fifth call
of the generator R(103,250), and
 found that  its severe problems seem to disappear.
Below I show that this five-call process is 
equivalent to making a single call of the four-tap generator 
R(50,103,200,250), and also discuss the 
inherent four-point correlations that that generator
possesses.  Coddington \cite{codd} and Vattulainen et al.\ \cite{vattu94} also 
utilized this decimation procedure.
From a speed point of view, however, it is clearly advantageous
to use the equivalent four-tap rule instead of having to make
multiple calls of a two-tap rule for each random number needed.

Recently, some lists of higher-order primitive
polynomials have appeared in the literature.
Those of Andr\'e et al.\ \cite{andre90} concern
relatively small offset values $p$ and 
have insufficient cycle lengths.
Note that these (and other) authors advocate
using many more feedback taps --- of the order
of $p/2$ --- which however would be impractical for the
large $p$ recommended here.  Some larger primitive pentanomials
have been given by Kurita and Matsumoto \cite{kurita} and more
recently by \v Zivkovi\' c \cite{zivkovic};
but none of these have been tested here.  (The present work
was carried out in 1992-94.)

The formulas for constructing new four-tap generators are
given in Section II, along with proofs.
In Section III, the correlations on smaller generators are 
found explicitly, and show that four-tap rules are 
vastly superior to two-tap rules in regards to three- and four-point
correlations, except for certain classes of
 four-tap
rules which have strong four-point correlations and probably should
 not be used.
In Section IV, a new test for random-number 
generators which  makes use of a kinetic self-avoiding
random walk related to percolation 
and the lattice Lorentz gas \cite{ruijcohen} is introduced.
While the two-tap and 
smaller four-tap generators badly fail the test, four-tap 
generators with moderately large offsets pass, and suggest that with
larger offsets, the errors should be nearly unmeasurable.
This test is evidently particularly sensitive 
to the type of asymmetric correlation that occurs these 
generators.  Some of our four-tap generators have also been
tested by Coddington \cite{codd} and 
Vattulainen et al.~\cite{vattu94}, who
confirmed the trends seen here.

\section{ Rules for four-tap generators.}

The taps $(a,b,c,\ldots)$ are chosen so that the 
corresponding polynomial $1 + z^a + z^b + z^c + \ldots $ is 
primitive over GF(2), guaranteeing 
that the cycle length will be the maximum possible value 
$2^p-1$, where $p=$ max($a,b,c,\ldots$)  \cite{gol67,peter72}.
Besides giving the maximum number of random numbers before
repeating, maximal rules have the advantage that they
can be initialized with any sequence (other than all zeros).
For two-tap rules, values of $a$ and $b$ can 
be found from extensive tables of primitive trinomials 
\cite{zb68,zier69,her92}.
Golomb has shown that higher-order polynomials can
be generated from trinomials by using 
a formal procedure based upon the concept of decimation \cite{gol67}.  

In $D$-decimation, every $D$-th term of a given sequence 
is selected to produce a new sequence.  The resulting sequence 
also satisfies a recursion like (\ref{eq:recur}), corresponding to a 
polynomial of order $p$, although the number of taps
is in general different.   
For some special cases of interest here,  I have found simple 
formulas which give four-tap rules 
directly.  Before deriving them, I first introduce 
the following 
alternate notation for the recursion (\ref{eq:recur}):  Let $[a, b, 
c, \ldots]$ indicate that the $x_n$ satisfy the relation
\begin{equation} 
x_{n-a}\ \oplus \ x_{n-b}\ \oplus \ x_{n-c}\ \oplus \ 
\ldots = 0 \qquad
\label{eq:corr}
\end{equation} 
for all $n$.  Thus, $[0,a,b]$ is an equivalent way to write 
(\ref{eq:recur}) for R$(a,b)$.  These relations satisfy some obvious 
properties:  If $[a,b,c\ldots]$ is satisfied on a given 
sequence, then $[a+k, b+k, c+k \ldots]$ will also be 
satisfied for any $k$ on that sequence (shift operation).
Furthermore, if 
both $[a, b, \ldots]$ and $[a', b', \ldots]$ are satisfied, 
then their union or sum $[a, a', b, b', \ldots ]$ will also 
be satisfied (addition property).  Finally, if an offset 
occurs twice in the list, then it can be eliminated, because 
$x_i \ \oplus \ x_i = 0$: $[a, b, b, c, \ldots] = [a, c, 
\ldots]$.  

Now, when a shift-register sequence is decimated by any 
power of 2, the original sequence is reproduced exactly, 
only shifted \cite{gol67}.  To prove this, consider the sequence 
generated 
by R(a,b)$ = [0,a,b]$.  By the shift property, $[a,2a,a+b]$ and $[b, 
a+b, 2b]$ are also satisfied on this sequence.  Adding these 
three relations together
yields $[0, a, a, b, b, a + b, a+b, 2a, 2b]$ = $[0, 2a, 2b]$,
which implies that every {\it other} 
term in the original sequence satisfies $[0,a,b]$.
  Thus, it follows that the 
original sequence and the two-decimated sequence must be identical.
Because the decimation wraps around the entire sequence,
which is odd in length, the 
decimated sequence is of the same maximal length as the 
original one.  This proof can be easily generalized for any 
(even) number of taps, and decimation by any power of 2.

When decimation by a number that is not a power of 2, a new 
sequence representing a different rule will, in general, be produced.
While in general the number of taps varies and may be large,
it turns out that 
four-tap rules always result when a two-tap rule R$(a,b)$ is 
decimated by $D=3,5$ and 7.  Those four-tap rules are given explicitly
by the following formulas:

\begin{equation}
{\rm R}(a,b) \times 3 = \cases{
{\rm R}(a, a/3, 2a/3, b) & $3 | a $ \hfill \qquad \qquad (3a) \cr 
{\rm R}(a, (2a+b)/3, (a+2b)/3, b) & $3 | (a-b)$ \hfill \qquad \qquad(3b) \cr } 
\label{eq:R3}
\end{equation}

\begin{equation}
{\rm R}(a,b) \times 5 = \cases{
{\rm R}(a, a/5, 4a/5, b) & $5 | a$ \hfill  \qquad\qquad(4a) \cr 
{\rm R}(a, (4a+b)/5,(a+4b)/5, b) & $5 | (a-b)$ \hfill \qquad \qquad(4b) \cr 
{\rm R}(a, (a+b)/5,2(a+b)/5, b) & $5 | (a+b)$ \hfill \qquad \qquad(4c) \cr 
{\rm R}(a, (3a+b)/5, (a+2b)/5, b) & $5 | (2a-b)$ \hfill  \qquad\qquad(4d) \cr } 
\label{eq:R5}
\end{equation}

\begin{equation}
{\rm R}(a,b) \times 7 = \cases{
{\rm R}(a, (a+b)/7, 3(a+b)/7, b) & $7|(a+b) $ \hfill \qquad \qquad(5a) \cr 
{\rm R}(a, (5a+b)/7, (a+3b)/7, b) & $7|(2a-b)$ \hfill \qquad \qquad(5b) \cr }
\label{eq:R7}
\end{equation}
where $D|a$ indicates that $a$ is divisible by $D$ (``$D$ divides $a$").
The remaining 
cases follow by switching $a$ and $b$ in the various formulas 
--- for example, when $2b-a$ is divisible by $D$, then $a$ 
and $b$ must be switched in (\ref{eq:R5}d) and (\ref{eq:R7}b).
Cases for $7|a$ and $7|(a-b)$ are not listed because these
cases do not occur among the primitive trinomials.

I deduced these decimation formulas by generating specific examples
using Golomb's methods \cite{gol67,gol69}, and finding
generalizations.  I then verified the formulas by 
application of the shift and  add properties given above.

For example, consider the case 
(\ref{eq:R3}a).  Shifting $[0,a,b]$ by $b, 2a, 2b$ and $a+b$ 
respectively yields the following five relations,

\begin{tabular}{lll}
&$ [0, a, b]$&original rule\\
&$ [b, a+b, 2b]$&rule shifted by $b$\\
&$ [2a, 3a, 2a+b]$&rule shifted by $2a$\\
&$ [2b, 2b+a, 3b] $ &rule shifted by $2b$\\
&$ [a+b, 2a+b, 2b+a] $&rule shifted by $a+b$\\
\end{tabular}

\noindent Summing these and canceling out common terms, one finds

$ [0, a, 2a, 3a, 3b] $

\noindent The final 
relationship (a five-point correlation) holds for any rule 
R($a,b$).  However, when $a$ is divisible by 3, then all five 
elements are divisible by 3, so it follows that the 3-decimated 
sequence satisfies $[0, a/3, 2a/3, 
a, b]$ or the rule R$(a/3, 2a/3, a, b)$ as given in (\ref{eq:R3}a). 

Likewise, for (\ref{eq:R3}b), we sum:

\begin{tabular}{lll}
&$ [0, 2a, 2b]$     &2-decimated rule \hfill \\
&$ [2a, 3a, 2a+b] $  &rule shifted by $2a$\hfill\\
&$ [2b, 2b+a, 3b]$   &rule shifted by $2b$\hfill\\
\end{tabular}

\noindent to find

 $ [0, 3a, 2a+b, a+2b, 3b] $

\noindent which implies (\ref{eq:R3}b) when $2a+b$ and $a+2b$ are both
divisible by 3, which occurs when $a-b$ is divisible by 3.  
For primitive trinomials, it is always true that either $a$, 
$b$, or $a-b$ is divisible by 3 \cite{gol67}, so (\ref{eq:R3}) contains 
all cases.  Proofs for 5- and 7-decimation are similar.  Note 
that decimations by more than 7 (and not a power of 2)
do not, in general, give 
four-tap rules but ones having many more taps.  In this 
regard, $D=3, 5,$ and 7 appear to be special cases.

Using the above formulas with $a$ and $b$ taken from existing tables of 
primitive trinomials \cite{zier69,her92}, numerous four-tap
generators can be found.  However, some 
of these generators will not be of maximal cycle length.
In order
that the cycle of the decimated sequence be the same as that 
of the original sequence, 
it is necessary that $D$ 
and $2^p-1$ have no common divisors, i.e., the g.c.d.$(D,2^p-
1)=1$.  This requirement is satisfied for 3-decimation when 
$p$ mod $2 \ne 0$, for 5-decimation when $p$ mod $4 \ne 0$, 
and for 7-decimation when $p$ mod $3 \ne 0$.  (On the other hand,
when these requirements are not satisfied, the 
cycle length is less than the maximum simply by a factor of 3, 5 or 
7, and is therefore still enormous when $p$ is large, so 
this consideration may not be so important.) 
An additional criterion for selecting which rules to decimate,
 concerning four-point correlations,
 will be discussed below.

\section{Correlations}

The relation $[a,b,c,\ldots]$ 
represents a correlation between the points $x_{n-a}$, 
$x_{n-b}$, $x_{n-c}$, $\ldots$. \
These are very strong correlations; for example, 
$[0,a,b]$, implies that if any two of $x_{n}, x_{n-a},$ and 
$x_{n- b}$ are known, the third is completely determined,
as mentioned above.  
The sequences generated by (\ref{eq:recur}) are literally laced with
 such correlations.  First of all, the 
basic correlation is given by the defining rule itself, R$(a, 
b, c, \ldots)$, in that $[0, a, b, c, \ldots]$ is satisfied 
for each $n$.  Then there is also a whole spectrum of three-point
correlations in the system:
By the so-called ``cycle and add" property 
\cite{gol67,comp87}, there exists an $s$ such that $[0,r,s]$ is satisfied 
for each value of $r = 1, 2, 3, \ldots 2^p-1$. \ The value of 
max($r,s$) is typically on the order of $2^{p/2}$ to $2^p$, when the 
defining rule is a pentanomial or higher.  However, when
the defining rule is a
trinomial R$(a,b)$, $s$ will be of the order $p$
for $r = a,\ b,\ 2a,\ 2b,\ 4a,\ 4b,\ $ etc.
These closely space three-point correlations interact to form  
numerous closely spaced four-point, five-point, and higher
correlations.

For most application, 
correlations involving the fewest number of points should be 
the most serious.  For example, if a kinetic random walk 
returns to the same region in space at steps $n$, $n-a$ and 
$n-b$ for some $n$, then its behavior would undoubtedly be 
affected by the three-point correlation [$0,a,b$] in the 
random-number sequence. Higher correlations would correspond 
to more coincidences in the motion of the walk and should 
therefore be less likely.  I will assume
that the reduction of three-point correlations is most 
important, followed by four-point correlations, and so on.

Using a four-tap rule R$(a,b,c,d)$ immediately eliminates the overriding 
three-point correlation [$0,a,b$] inherent in a two-tap rule 
R($a,b$), and the remaining three-point correlations are
widely spaced as mentioned above.
The four-point correlations of a four-tap rule care
also generally widely spaced.  An exception
occurs when the four-tap rule follows from a $D$-decimation
of a two-tap rule R$(a,b)$ and
$a$, $b$, or $a-b$ is divisible by $D$.  In this case,
the correlation offsets are small and can be derived explicitly.
For example, the 3-decimation of R($a,b$) yields $[0, 
a/3, 2a/3, a, b]$ according to (\ref{eq:R3}a).
By shifting this five-point correlation and adding,
one finds the four-point correlation 
\begin{equation}
 [0, a/3, 2a/3, a, b] + [a/3, 2a/3, a, 4a/3, a/3 + b]= [0, 
4a/3, b, a/3+b]
\label{eq:4pt}
\end{equation}
The spread of this correlation 
is of the order of $p$, not $2^p$. 
Such a four-point correlation in R$(38,89) \times 3 = $R$(38,55,72,89)$  
(where $89-38$ is divisible by 3)
was noted in \cite{shchur}.
A similar result holds for the 5-decimation rules (\ref{eq:R5}a,b).  
Therefore, to avoid these relatively close four-point correlations,
all 3-decimations (\ref{eq:R3}) and the 5-decimations 
(\ref{eq:R5}a,b) should not be used, and 
will not be considered further below, except for
the rule  R$(103,250) \times 5$ = R$(50,103,200,250)$
which was considered in$\cite{ferr92}$.
Here, 250 is divisible by 5, and as a consequence the sequence obeys 
the relatively closely spaced
four-point correlation [0, 309, 359, 800].

For generators produced by other rules, it appears that the 
correlations can only be found by 
a search procedure, in which a sequence of 
bits is generated, and different correlations are checked 
until the sequence is matched.  To make this feasible for
larger $p$, I made a list of up to $2^{21}$
32-bit sub-sequences,
and sorted them with keys pointing to their location in the sequence,
in order to be able to quickly find if a sequence generated by a trial correlation
occurs.  Details will be presented elsewhere.
This procedure turned out to be
practical for finding three- and four-point correlations
for $p$ up to about 50.
 
Some representative results from this search are given below.
Each line shows
respectively the way the rule was generated from the two-tap 
rules of \cite{zier69}, the equivalent four-tap rule 
from (\ref{eq:R5}) or 
(\ref{eq:R7}) (which also represents the smallest five-point correlation 
$[0,a,b,c,d]$ in the sequence), and the smallest four- and three-point
correlations found by our search.  
These results are:
\begin{mathletters}
\begin{eqnarray} 
&&{\rm R}(5,17)\times 7 = {\rm R}(5,6,8,17) = [0,77,79,101] = 
[0,67,83] \\
&&{\rm R}(5,23)\times 7 = {\rm R}(4,5,12,23) = [0,13,50,421] = 
[0,1153,4933] \\
&&{\rm R}(3,31)\times 5 = {\rm R}(3,8,13,31) = [0,87,199,397] = 
[0,30189,34284] \\
&&{\rm R}(6,31)\times 7 = {\rm R}(6,7,23,31) = [0,40,623,2216] 
= [0,14487,101088] \\
&&{\rm R}(8,39)\times 7 = {\rm R}(8,9,29,39) = 
[0,111,1072,7006] = [0,172074,758257] \\
&&{\rm R}(3,41)\times 7 = {\rm R}(3,8,18,41) = 
[0,4280,6131,8713] = [0,351102,1716109] \\
&&{\rm R}(20,47)\times 7 = {\rm R}(20,21,23,47) = 
[0,33579,138448,150900] = [0,8474125,11136544] \\
&&{\rm R}(21,47)\times 5 = {\rm R}(21,22,23,47) = 
[0,63608,148485,156350]  = [0,11941097,13215912]
\end{eqnarray} 
\end{mathletters}
Thus, for example, the four-tap rule R(5,6,8,17) generates a 
series that has the three-point correlation [0, 67, 83], 
four-point correlation [0,77,79,101], as well as the
inherent five-point 
correlation [0,5,6,8,17] (not shown explicitly).
Note that the two-tap rule 
R(67,83) corresponding to this three-tap correlation can only
be used to generate the sequence produced by  R$(5,6,8,17)$
if it is started up correctly with the 83 bits from the latter's
sequence, because the sequence generated by R$(5,6,8,17)$ is
only one of many cycles of the non-maximal rule R$(67,83)$.  
Therefore, the correlations in brackets, such as 
[0, 67, 83],  should not be interpreted as suggested rules
for random-number generators.

The above results clearly show that the 
separation in the three-and four-point correlations increases 
rapidly as $p$ increases.  In fact,
the extent of the smallest three-point correlation grows
roughly as $2^{p/2}$, and the extent of the smallest
four-point correlations as $2^{p/3}$.
Clearly, for larger $p$, such correlations will be 
irrelevant, and the most important 
correlations in four-tap rules will be the five-point
ones generated by the rule itself.

Additional maximal length rules can be generated
by Golomb's method of repeated 3-decimation \cite{gol69}.
(For some cases of $p$, 
repeated 3-decimation of a single maximal-length rule  
yields the complete cycle of all possible maximal-length rules.) 
For comparison, I have studied the behavior of some of 
these other rules.  I found that, for a given $p$, the three-
and four-point correlations have roughly the same separation 
as found for the rules that follow from simple 5- and
7-decimation.
For example, for the four-tap rule 
R$(23,27,40,41)$, found by successively
3-decimating R$(3,41)$ $107005025$ times 
--- equivalent to decimating once by $3^{107005025}$ mod 
$({2^{41}-1}) = 1962142349662$ --- I find 
\begin{eqnarray}
{\rm R}(3,41) && \times 1962142349662 = 
{\rm R}(23,27,40,41) \nonumber \\
&&= [0,20573,22443,25575] = 
[0,429959,1013792] \label{eq:8}
\end{eqnarray}
which may be compared with (7f).  Six-tap rules with 
$p = 41$ were found to possess similar three- and four-point
correlations.

Thus, for useful generators,
we turn to rules with larger $p$.
Following are some larger four-tap rules generated by (\ref{eq:R5}b,c)
and (\ref{eq:R7}): 

\begin{mathletters}
\begin{eqnarray}
&&{\rm R}(38,89) \times 5 = {\rm R}(33,38,61,89) \\
&&{\rm R}(11,218) \times 7 = {\rm R}(11,39,95,218) \\
&&{\rm R}(216,1279)\times 5 = {\rm R}(216,299,598,1279) \\
&&{\rm R}(216,1279)\times 7 = {\rm R}(216,337,579,1279) \\
&&{\rm R}(471,9689) \times 5 = {\rm R}(471,2032,4064,9689) \\
&&{\rm R}(471,9689) \times 7 = {\rm R}(471,1586,6988,9689) \\
&&{\rm R}(33912,132049)\times 5 = {\rm R}(33912,46757,59602,132049) \\
&&{\rm R}(33912,132049)\times 7 = {\rm R}(33912,43087,61437,132049) 
\end{eqnarray}
\end{mathletters}
The three- and four-point correlations for these rules are 
undoubtedly much larger than can be found by my search program.  To 
assess the quality of these generators, I turn to a test based 
upon a problem from percolation theory.

\section{Test on random walk problem}

The test I use is shown in Fig.~1. \  A walker starts at the 
lower left-hand corner of a square lattice, and heads in the 
diagonal direction toward the opposite corner.  At each step 
it turns at a right angle either clockwise or 
counter-clockwise.  When it encounters a site it had never 
visited before, the walker chooses which direction to turn 
with a 50-50 probability, while at a site that has been 
previously visited, it always turns so as not to retrace its 
path  (a so-called kinetic self-avoiding 
trail on a square lattice).  The boundary of the 
lattice is a square; the lower and left-hand sides 
are reflecting, while the upper and right-hand side are 
adsorbing.  Clearly, by the perfect symmetry of the problem, 
the walker should first reach either the top or the right-hand
sides with equal probability.  We shall see, however, 
that not all these random-number generators yield this simple 
result.

It turns out that this walk is precisely the kinetic self-avoiding 
walk that generates the hull of a bond percolation 
cluster at criticality.
The lattice vertices visited by the walk are located at the centers of 
the bonds, and the two choices correspond to placing either a bond 
on the lattice or one on the dual lattice across that vertex point. The 
1/2 probability of reaching the upper side first corresponds 
to a spanning or crossing probability of exactly 1/2 for 
this system \cite{cardy,langlands,ziff92}.
The walk is also identical to a 
lattice-Lorentz gas introduced by Ruijgrok and Cohen 
\cite{ruijcohen} with randomly oriented mirrors, to motion 
through a system of rotators as introduced by Gunn and Ortuno 
\cite{gunn}, and to paths on the random tiling of 
Roux et al.~\cite{roux}.   Note that this test is an actual
algorithm that has been used
in percolation studies \cite{ziff92,ziff96,hovi}; it is not
a ``cooked-up" problem designed specifically to reveal flaws in 
a specific random-number generator.

Using this procedure, I tested a variety of generators,
including the two-tap generators R(11,218), R(103,250),
R(216,1279), R(576,3217), and R(471,9689),
the four-tap generators R(20,27,34,41), R(3,26,40,41), R(1,15,38,41),
R(1,3,4,64), R(33,38,61,89),
R(11,39,95,218), R(50,103,200,250), R(216,337,579,1279), and
R(471,1586,6988,9689), and the six-tap generators
(determined through successive 3-decimation \cite{gol69})
R(1,5,8,30,35,41), R(5,14,20,36,37,41), and R(18,36,37,71,89,124).
Between 100,000 and 2,000,000 
trials were simulated with each generator,
yielding an error of about $\pm 0.001$.
The lattice was of size $4096\times$4096, and intermediate 
results for squares of side $L= 64, 128,192, \ldots, 4032$
were also recorded. 
Figs.~2 and 3  show the fraction of walks that first arrived 
at the upper boundary in each of these runs as a function
of $L$.  Clearly, some 
generators are very bad; for example, with the notorious R(103,250),
 the top of the $4096 \times 4096$ square was reached only 32\% 
of the time! 
This error clearly cannot be statistical in origin;
in fact, it is about 180 times the standard 
deviation $\sigma = 0.001$.  All of the smaller two-tap 
generators are clearly quite poor, but even the largest one
with $p=9689$  is barely within two standard 
deviations at $L=4096$.  

On the other hand, the four-tap generator with $p=89$ begins to 
show deviations only at the largest $L$, and the generator 
with $p=218$ shows no visible deviations at all in this work.
(However, in more recent tests of $10^8$ runs on a lattice of size 256 $\times$ 256,
I found some error creeping in for R(11,39,95,218),
with the crossing probability at $L=256$ given by
$0.50030 \pm 0.00005$ \cite{ziff96}.) \ 
Clearly, as $p$ is increased, more random numbers need to
be generated before the errors can be seen.  For four-tap rules
with $p$ larger than about 500, it appears that deviations in this test
would be nearly impossible to uncover with present-day computers.

There are a number of interesting and puzzling aspects
of these results.
Evidently, two-tap generators were give low results, four-tap
generators give high results, and six-tap ones again give low
ones.  The supposedly
bad generator R(50,103,200,250), with its strong four-point
correlations mentioned above, actually yields excellent results.
Finally, the generators R(3,26,40,41) and R(1,15,38,41)
are mirrors of each other, and so have identical (but mirrored)
correlations of all points, and yet give noticeably different 
behavior.  The explanation of these intriguing properties is a subject
for future research.  One might also investigate whether the 
choice to grow a new hull immediately after the previous
has completed, without any gap the random number sequence,
has any bearing on the results.

Note that the plots in Figs.~2 and 3 are nearly, but not
quite, linear.  In fact, one can argue that the behavior must
grow with a power of $L$ that is less than or equal to 7/8.  
For, say that the error grows as $L^x$ with increasing $L$.  This error
will first be discernible when the number of runs $N_{runs}$
satisfies $N_{runs}^{-1/2} \sim L^x$ or $N_{runs} \sim L^{-2x}$.
The number of random numbers generated per run grows as
$L^{7/4}$, where 7/4 is the fractal dimension of the
hull.  Thus, the total number of random numbers generated grows as 
$\sim L^{7/4 - 2x}$.  Now, the exponent in the latter expression
 cannot be negative,
since that would imply that going to an infinite system would allow
the error to be found with no work.
So we deduce
$x \le 0.875$.   Numerically, a value of about $x = 0.7$ seems to give
the best fit to the data in Fig.~2.   That $x$ is less that $0.875$ implies
that doing more runs on a smaller lattice, rather than fewer runs
on a larger lattice, is actually a more efficient
way to uncover the errors in these generators, assuming the
same power-law behavior of the error holds for small $L$.

Because this test is completely symmetric, the errors seen here
highlight the fundamental asymmetry of the GFSR generator.
Indeed, the basic exclusive-or 
operation has an asymmetry to it, as two 0's or two 
1's both result in a 0.  For a correlation or generator 
$[0,a,b]$, the three points $x_n$, $x_{n-a}$, and $x_{n-b}$ 
can have only the values (0,0,0) and (0,1,1) (and 
permutations) which is clearly not symmetric.  (This
asymmetry is not in the total abundance of 0's and 1's, 
which are equally probable, but in their correlations.) \  
Another way of demonstrating this asymmetry is to note that
changing 1's to 0's and 0's to 1's in the initial seed 
sequence  does not result in the complementary sequence being 
generated.  That is, complementary sub-sequences are not equally likely.

We also carried out test
 with 31- and 48-bit
linear congruence generators, and no
errors were found.  Evidently, these generators have a symmetry
such that complementary sequences are generated with equal
probability, which leads to a probability of reaching the top
of exactly 1/2. 
This result underscores the proviso that the test 
used here is not relevant for all random-number
generators --- as, indeed, no test is.

\section{Conclusions}

Clearly, all GFSR random-number generators will 
eventually show some detectable errors if
a sufficiently long run is made.  However, when
the four-tap generators with $p$ greater than about 500 
is used, the amount of computer time needed to uncover those
errors will be prohibitive.  Three- and four-point correlations
of these generators are projected to be enormously spread
apart.  Thus, such large four-tap generators appear
useful as a practical, high quality
pseudorandom-number generator.

Two-tap generators, in contrast, do not pass the test 
carried out here, except perhaps those with the
largest tap offsets.  Thus, 
for critical applications, it appears that all two-tap  
generators, not just R(103,250), should be excluded.

 If a problem is sensitive to the built-in
five-point correlations of a four-tap generator, then a higher 
number of taps should be used.  For this, the combination 
generator discussed by Compagner \cite{comp91,comp91a} is useful.

In spite of their known problems, 
there are many reasons that GFSR random-number
generators remain of interest.
In contrast to some combination generators, they are clean and
well-characterized; 
a large body of fundamental theory on their properties has been 
produced (i.e., \cite{nied92}). 
Even with four taps, they remain 
fast and easy to program.  Each bit is entirely
independent, which is not the case for linear congruence
generators or ``lagged-Fibonacci" generators with
addition or multiplication.
Although they require storing a long list to exhibit good behavior, the memory
requirements are not a problem for present-day computers.  

Over the last 10 years, we have carried out numerous
extensive simulations on a variety of problems in 
percolation and interacting particle models
using the four-tap generators derived here.
Our earlier work (i.e., \cite{ziffstell}) made use of R$(157,314,471,9689)$
which derives from R$(471,9689) \times 3$; more recently
(i.e., \cite{ziff92,ziff96,suding}) we switched
to the 7-decimation generator (9f) given 
above, because of the inherent four-point correlations
in a 3-decimation rule as discussed in this paper
(although we never observed any problem with the former,
presumably because of its large $p$).
In all this work, in which we often made checks
with exact results when available, we
never found any indication of error.   In a recent paper determining
the bond percolation threshold for the Kagom\'e lattice \cite{suding},
we also checked the results of using R(471,1586,6988,9689)
against runs using a 64-bit congruential
generator, as well as the 3-decimation of R(471,1586,6988,9689)
(thus equivalent to R(471,9689) $\times$ 21),
and found complete consistency throughout.

In closing, I give an explicit example to of the
generator, written in a single line of the C programming language.
It makes use of the {\tt define} 
statement, which results in in-text substitution during the 
pre-compiling stage, so that no time is lost in a 
function call:
\begin{verbatim}
#define RandomInteger (++nd, ra[nd & M] = ra[(nd-A) & M] \
   ^ ra[(nd-B) & M] ^ ra[(nd-C) & M] ^ ra[(nd-D) & M]) 
\end{verbatim}
\noindent  The generator is called simply as follows
\begin{verbatim}
   if (RandomInteger < prob) ... 
\end{verbatim}

\noindent where, for rule (9f) for example,
{\tt A=471, B=1586, C=6988, D=9689;} and 
{\tt M = 16383} (defined as constants), {\tt ra} is an integer
array over {\tt 0..M} that is
typically initialized using a standard congruential random-number,
{\tt nd} is its index (an integer),
{\tt \&} is the bitwise ``and" 
operation, and {\tt \^ \ }is the bitwise ``xor" operation.
``Anding" with {\tt M} effectively causes the numbers to 
cycle endlessly 
around the list, when {\tt M+1} is chosen to be
a power of two as above.  The list in this example requires
64 kilobytes of memory ($16384 \cdot 4$),
if 32-bit (4-byte) integers  are used.
\  Here, {\tt prob}
is the probability of the event occurring, converted to an integer
in the range of 0 to the maximum integer.  A floating-point
number can also be produced by dividing {\tt RandomInteger} by the
maximum integer (which depends upon the
number of bits in the generator), but this added step 
consumes additional time.  Using the above program,
an HP 9000/780 workstation computer generates a
random number in about 50 nanoseconds,
or one billion ($10^9$) in less than a minute.

Acknowledgments:
I  acknowledge useful correspondence with H. Fredricksen 
and S. Golomb a number of years ago, and many helpful 
questions and comments from P. Grassberger in an extended 
electronic mail interchange.  I also thank P. Coddington 
and I. Vattulainen, T. Ala-Nissila, and K. Kankaala
for testing some of these four-tap generators and sharing their 
results with me prior to publication, as well as  A. Compagner for
sending me preprints of his work over the years.
 This material is based upon work
supported by the US National Science Foundation under
grant no. DMR-9520700.

\begin{figure}
\caption{The random walk algorithm used to test the random numbers,
shown for a system of size $L=4$.  
The test is whether the walker, which turns $90^\circ$ to the left
or right with equal probability at each newly visited site, first
reaches the right or top with equal probability.  The left and bottom sides are
reflecting. }
\end{figure}

\begin{figure}
\caption{Plot of the probability of the walk reaching the top of an
$L \times L$ system, vs.\ $L$, showing large deviations from the
expected value of 1/2 for many of the generators. }
\end{figure}

\begin{figure}
\caption{Central portion of Fig.~1 expanded vertically. }
\end{figure}

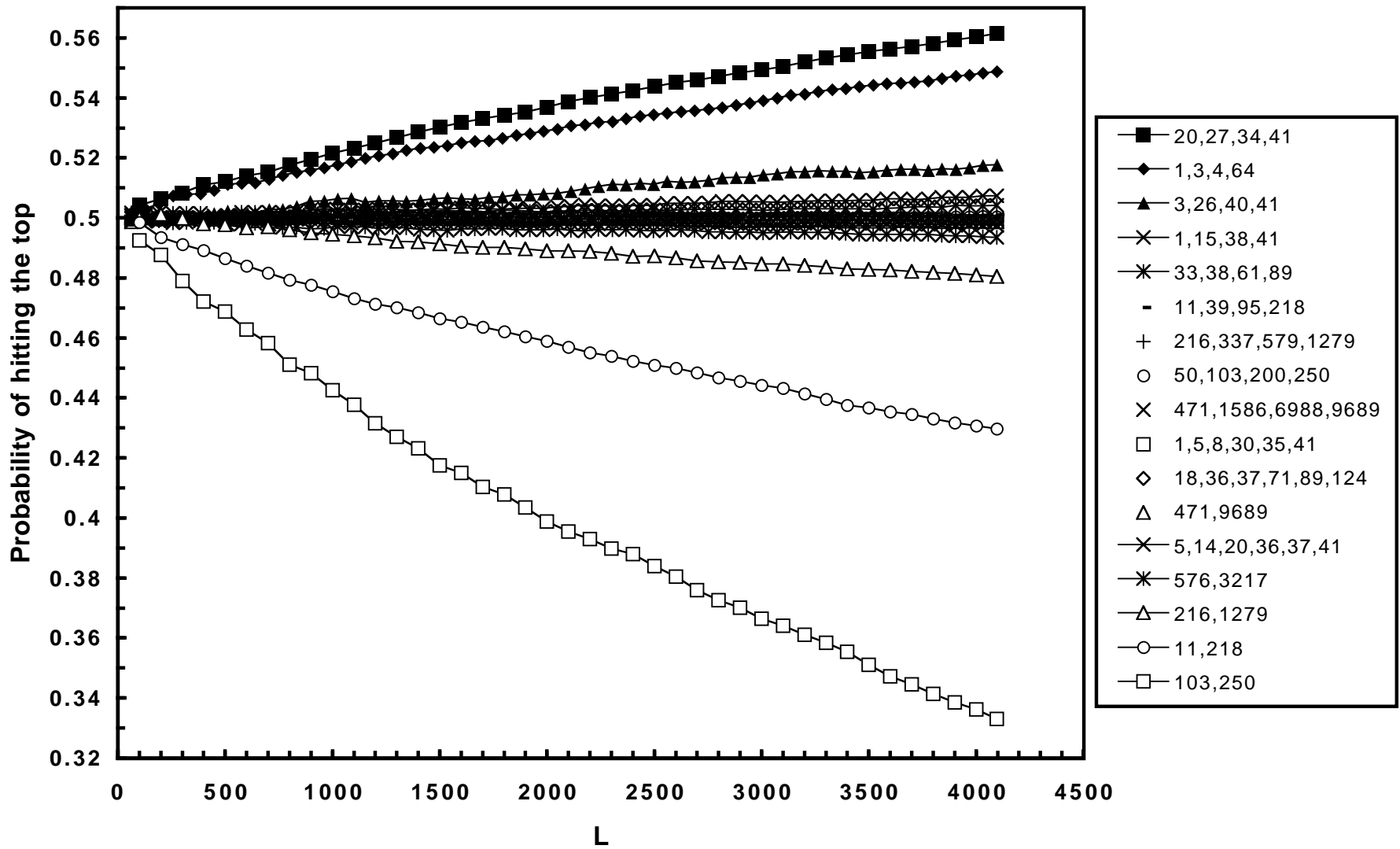
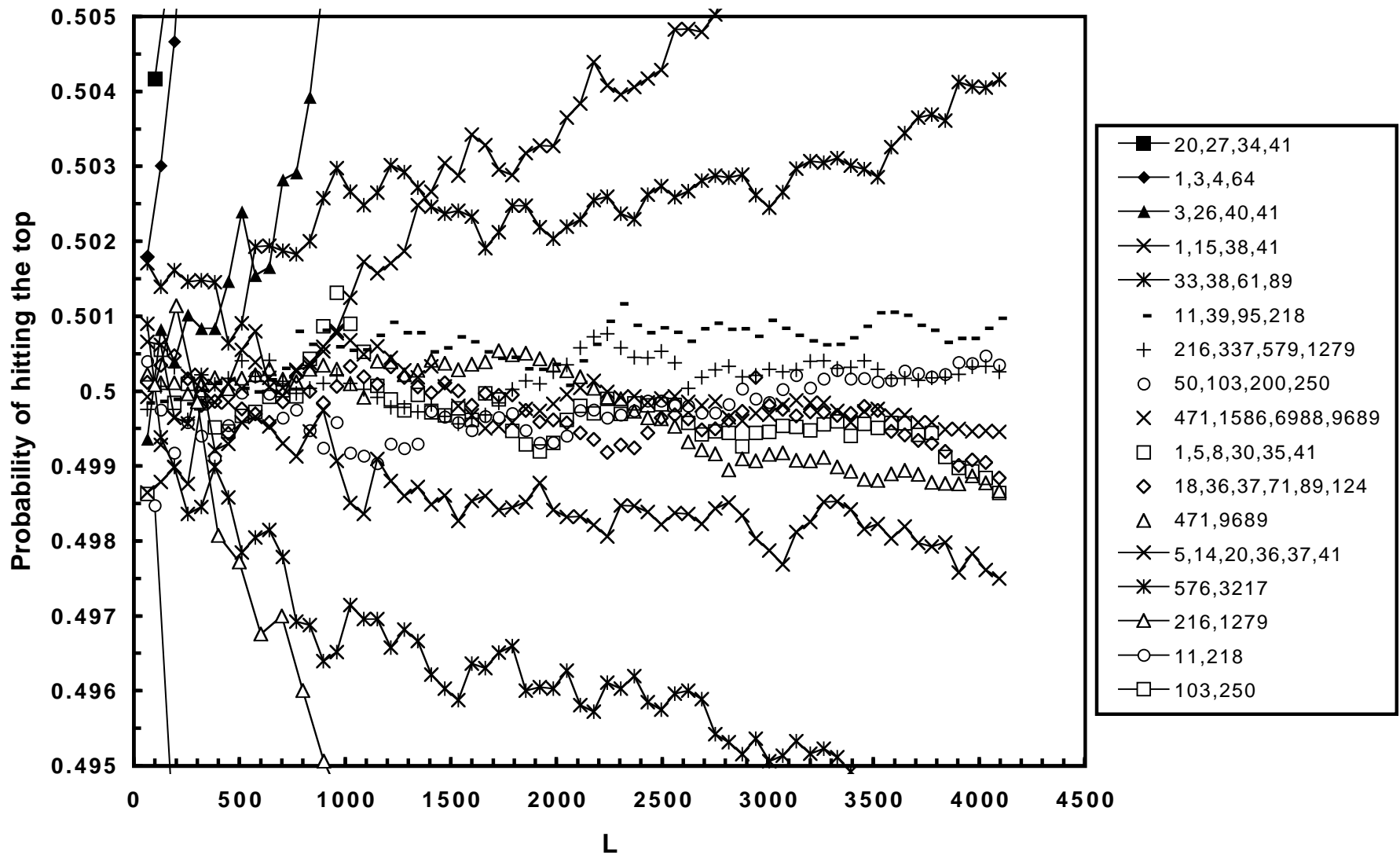
\end{document}